\documentclass[reprint, amsmath, twocolumn, secnumarabic, amssymb, nobibnotes, aps, prapplied, superscriptaddress]{revtex4-2}

\newcommand{\env}[1]{\texttt{#1}}
\usepackage{graphics}
\usepackage{graphicx}
\usepackage{bm}
\usepackage{amsmath}
\usepackage{verbatim}
\usepackage{hyperref}
\usepackage{threeparttable}
\usepackage{courier}

\begin{document}

\title{Qubit Energy Tuner Based on Single Flux Quantum Circuits}%

\author{Xiao Geng}%
\email[Email: ]{gengx19@mails.tsinghua.edn.cn}
\affiliation{Laboratory of Superconducting Quantum Information Processing, School of Integrated Circuits, Tsinghua University, Beijing 100084, China}
\affiliation{Beijing National Research Center for Information Science and Technology, Beijing 100084, China}

\author{Rutian Huang}
\affiliation{Laboratory of Superconducting Quantum Information Processing, School of Integrated Circuits, Tsinghua University, Beijing 100084, China}
\affiliation{Beijing National Research Center for Information Science and Technology, Beijing 100084, China}

\author{Yongcheng He}
\affiliation{Laboratory of Superconducting Quantum Information Processing, School of Integrated Circuits, Tsinghua University, Beijing 100084, China}
\affiliation{Beijing National Research Center for Information Science and Technology, Beijing 100084, China}

\author{Kaiyong He}
\affiliation{Laboratory of Superconducting Quantum Information Processing, School of Integrated Circuits, Tsinghua University, Beijing 100084, China}
\affiliation{Beijing National Research Center for Information Science and Technology, Beijing 100084, China}

\author{Genting Dai}
\affiliation{Laboratory of Superconducting Quantum Information Processing, School of Integrated Circuits, Tsinghua University, Beijing 100084, China}
\affiliation{Beijing National Research Center for Information Science and Technology, Beijing 100084, China}

\author{Liangliang Yang}
\affiliation{Laboratory of Superconducting Quantum Information Processing, School of Integrated Circuits, Tsinghua University, Beijing 100084, China}
\affiliation{Beijing National Research Center for Information Science and Technology, Beijing 100084, China}

\author{Xinyu Wu}
\affiliation{Laboratory of Superconducting Quantum Information Processing, School of Integrated Circuits, Tsinghua University, Beijing 100084, China}
\affiliation{Beijing National Research Center for Information Science and Technology, Beijing 100084, China}

\author{Qing Yu}
\affiliation{Laboratory of Superconducting Quantum Information Processing, School of Integrated Circuits, Tsinghua University, Beijing 100084, China}
\affiliation{Beijing National Research Center for Information Science and Technology, Beijing 100084, China}

\author{Mingjun Cheng}
\affiliation{Laboratory of Superconducting Quantum Information Processing, School of Integrated Circuits, Tsinghua University, Beijing 100084, China}
\affiliation{Beijing National Research Center for Information Science and Technology, Beijing 100084, China}

\author{Guodong Chen}
\affiliation{Laboratory of Superconducting Quantum Information Processing, School of Integrated Circuits, Tsinghua University, Beijing 100084, China}
\affiliation{Beijing National Research Center for Information Science and Technology, Beijing 100084, China}

\author{Jianshe Liu}
\affiliation{Laboratory of Superconducting Quantum Information Processing, School of Integrated Circuits, Tsinghua University, Beijing 100084, China}
\affiliation{Beijing National Research Center for Information Science and Technology, Beijing 100084, China}

\author{Wei Chen}
\email[Email: ]{weichen@mail.tsinghua.edu.cn}
\affiliation{Laboratory of Superconducting Quantum Information Processing, School of Integrated Circuits, Tsinghua University, Beijing 100084, China}
\affiliation{Beijing National Research Center for Information Science and Technology, Beijing 100084, China}
\affiliation{Beijing Innovation Center for Future Chips, Tsinghua University, Beijing 100084, China}
\date{\today}

\begin{abstract}
\par
A device called qubit energy tuner (QET) based on single flux quantum (SFQ) circuits is proposed for Z control of superconducting qubits. Created from the improvement of flux digital-to-analog converters (flux DACs), a QET is able to set the energy levels or the frequencies of qubits, especially flux-tunable transmons, and perform gate operations requiring Z control. The circuit structure of QET is elucidated, which consists of an inductor loop and flux bias units for coarse tuning or fine tuning. The key feature of a QET is analyzed to understand how SFQ pulses change the inductor loop current, which provides external flux for qubits. To verify the functionality of the QET, three simulations are carried out. The first one verifies the responses of the inductor loop current to SFQ pulses. The results show that there is about 4.2\% relative deviation between analytical solutions of the inductor loop current and the solutions from WRSpice time-domain simulation. The second and the third simulations with QuTip show how a Z gate and an iSWAP gate can be performed by this QET, respectively, with corresponding fidelities 99.99884\% and 99.93906\% for only once gate operation to specific initial states. These simulations indicate that the SFQ-based QET could act as an efficient component of SFQ-based quantum-classical interfaces for digital Z control of large-scale superconducting quantum computers.
\end{abstract}

\maketitle

\section{Introduction}
\par
Josephson qubits with gate and measurement fidelities over the threshold of fault-tolerant quantum computing are an attractive candidate for manufacturing scalable quantum computers. As a traditional way for qubit control and readout, microwave electronics successed in obtaining gate fidelities beyond 99.9\% \cite{barends2014superconducting} and realizing quantum supremacy \cite{arute2019quantum}. Because of quantitative restrictions to input and output  ports of quantum processor and cryogenic transmission lines, the bottleneck of interconnection comes to be significant when the number of qubits increase beyond a thousand. To overcome the bottleneck, it is desirable to introduce single flux quantum (SFQ) digital logic circuits \cite{likharev1991rsfq} for control and readout \cite{mcdermott2018quantum}. Digital coherent XY control based on SFQ pulses to transmon qubits was proposed \cite{mcdermott2014accurate} and the fidelities of digital single-qubit gates were measured to be about 95\% \cite{leonard2019digital}. Methods of optimization to SFQ pulse sequences for single-qubit gates \cite{liebermann2016optimal, li2019hardware} and two-qubit gates like cross-resonance gates and controlled phase (CZ) gates \cite{jokar2022digiq, dalgaard2020global, jokar2021practical} were also studied.
\par
To control qubits flexibly, SFQ-based devices for Z control has become a frontier requiring more research. In 2018, McDermott \textit{et al.} \cite{mcdermott2018quantum} proposed an SFQ-based coprocessor working at 3 K for control and measurement of a quantum processor requiring SFQ-based flux digital-to-analog converters (flux DACs) \cite{johnson2010scalable} for Z control, which is really an inspiring idea for creating a scalable superconducting quantum processor. Recently, Mohammad \textit{et al.} \cite{jokar2022digiq} proposed an SFQ-based digital controller called \textit{DigiQ} for superconducting qubits, in which the Z control of qubits are performed with bias currents generated by an array of SFQ/DCs. These SFQ/DCs in \textit{DigiQ} are placed at the 4 K plate of the dilution refrigerator, so bias currents need to be transmitted in superconducting microstrip flex lines to 10 mK plate where the quantum processor works, which is similar to Ref.\onlinecite{mcdermott2018quantum}.
\par
In order to promote integration further, SFQ logic circuits for control and measurement of qubits should be integrated with quantum processor with 3D integration technologies \cite{rosenberg20173d, rosenberg2020solid} in the future. Therefore, SFQ-based devices for on-chip Z control need to be researched and designed with circuits as simple as possible for scalable quantum processors. These simple devices should be able to convert SFQ pulse signals to flux signals, just like flux DACs. The circuits for Z control in \textit{DigiQ} may be a little more complicated than flux DACs, so intuitively flux DACs can be considered as the first choice for developing SFQ-based devices for Z control. However, with a single flux DAC defined in Ref.\onlinecite{johnson2010scalable}, it is difficult to provide flux bias and simultaneously complete a Z-control gate with high precision due to the following reasons. (i)The resetting of a flux DAC at the end of the gate can eliminate not only the flux performing the gate but also the bias flux setting the idle frequency of the controlled qubit. (ii)The resetting is accomplished by applying $\Phi_0/2$ (half of a flux quantum) to the two-junction reset SQUID loop, which may require another flux DAC, an SFQ/DC or a pin of the coprocessor for an external current source. This finally increases the physical footprint and complexity of the coprocessor, or aggravates the situation of interconnection bottleneck.
\par
Here, we propose a new SFQ-based device for Z control, which is created from the improvement of flux DACs. Because its basic function is to tune the energy levels of qubits, it is called qubit energy tuner (QET). With flux provided by a QET, the energy levels or the frequency of a flux-tunable transmon qubit can be set to specific levels. At the same time, it can also perform gate operations which need flux bias, such as a Z gate or an iSWAP gate. After a gate operation, the QET can tune the frequency of the qubit back to its idle frequency.
\par
In this article, the circuit structure of QETs is first described in Section II. The key feature of a specific QET is analyzed and an formula is obtained for calculating its inductor loop current providing flux. Next, the ideal Z control method with square-wave-like currents for flux-tunable tranmson is discribed in Section III. Then in Section IV, simulations are done for presenting how the inductor loop current of a QET that provides flux is changed by SFQ pulses and how a QET performs a Z gate and an iSWAP gate. In Section V, the work in this article is concluded. The challenges and opportunities about the QET in the future are discussed.

\section{Structure of Qubit Energy Tuner}

\begin{figure}
\begin{center}
\includegraphics[scale=0.85]{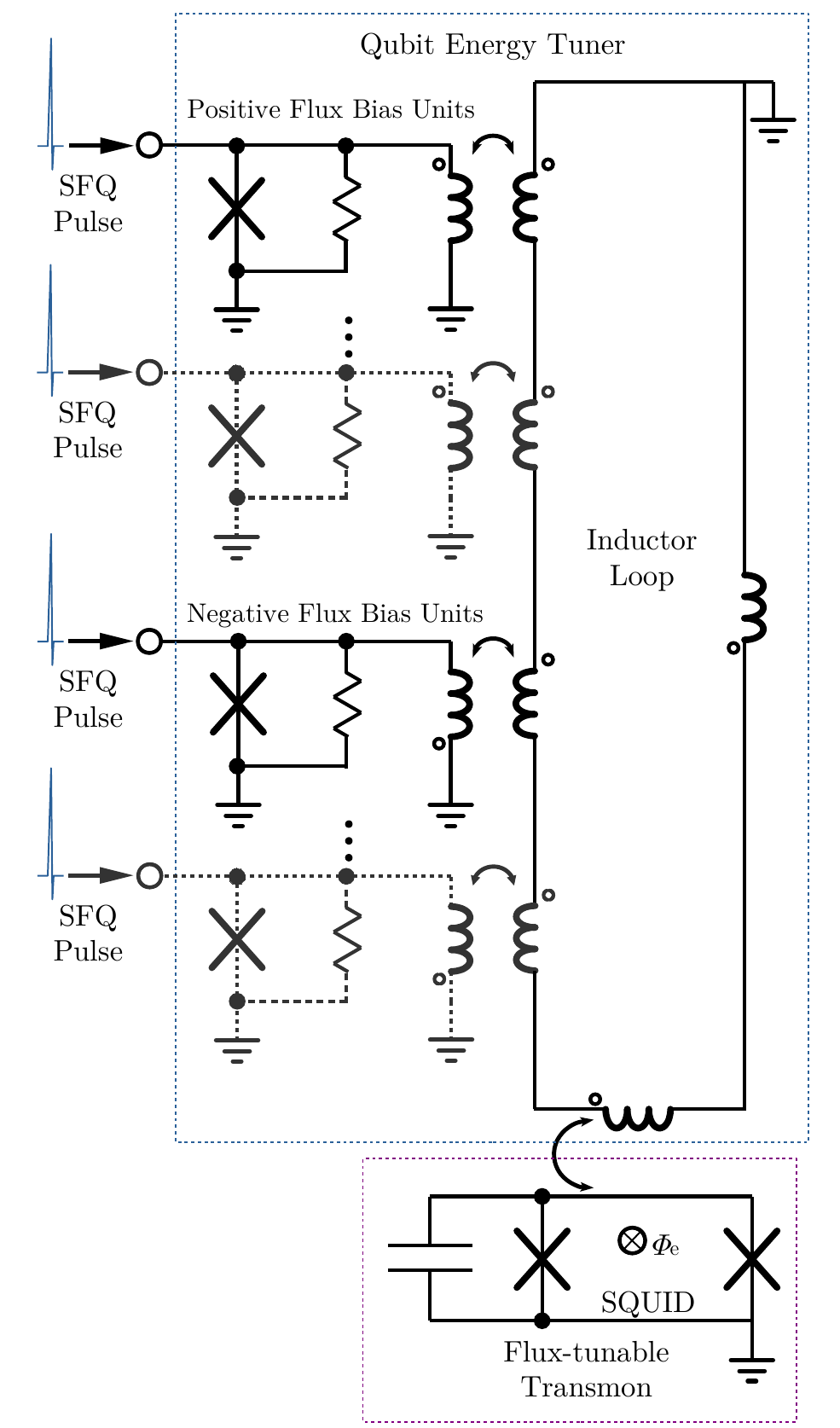}
\caption{\label{stuqet} Structure of a qubit energy tuner coupled with a flux-tunable transmon.}
\end{center}
\end{figure}

\par
A qubit energy tuner (QET) contains an inductor loop and some flux bias units, positive or negative, as can be seen in FIG.~\ref{stuqet}. The inductor loop is weakly coupled to the SQUID of a flux-tunable transmon, providing flux for tuning the energy levels of the transmon. A flux bias unit includes a Josephson junction shunted with an inductor, which is coupled to the inductor loop. The Josephson junction can be made of an intrinsic Josephson junction in parallel with a resistor to be an overdamped Josephson junction. The node connected to the Josephson junction and the inductor is treated as an input port of QET for SFQ pulse signal. After recieving an SFQ pulse, a positive flux bias unit increase the external flux through the SQUID by a specific amount while a negative flux bias unit increase it in the opposite direction or decrease it by the same specific amount. This is realized by making the direction of dotted terminals of positive flux bias units the same as that of the corresponding inductor in the inductor loop but making the direction of dotted terminals of negative flux bias units opposite to that of the corresponding inductor in the inductor loop.
\par
A QET should have at least a pair of flux bias units, in which one is positive and the other is negative. In order to tune the energy levels of a transmon more precisely, QET can be designed to have two or more pairs of flux bias units, among which some pairs are used for coarse tuning and others are used for fine tuning. Inductors in different flux bias units can also be coupled for optimization of circuit performance.
The inspiration about the qubit energy tuner is from the design of the flux DAC proposed by Ref.~\onlinecite{johnson2010scalable} and Ref.~\onlinecite{bunyk2014architectural}. Therefore, its circuits have similar but simpler structures compared with those of flux DACs.

\begin{figure}
\begin{center}
\includegraphics[scale=0.85]{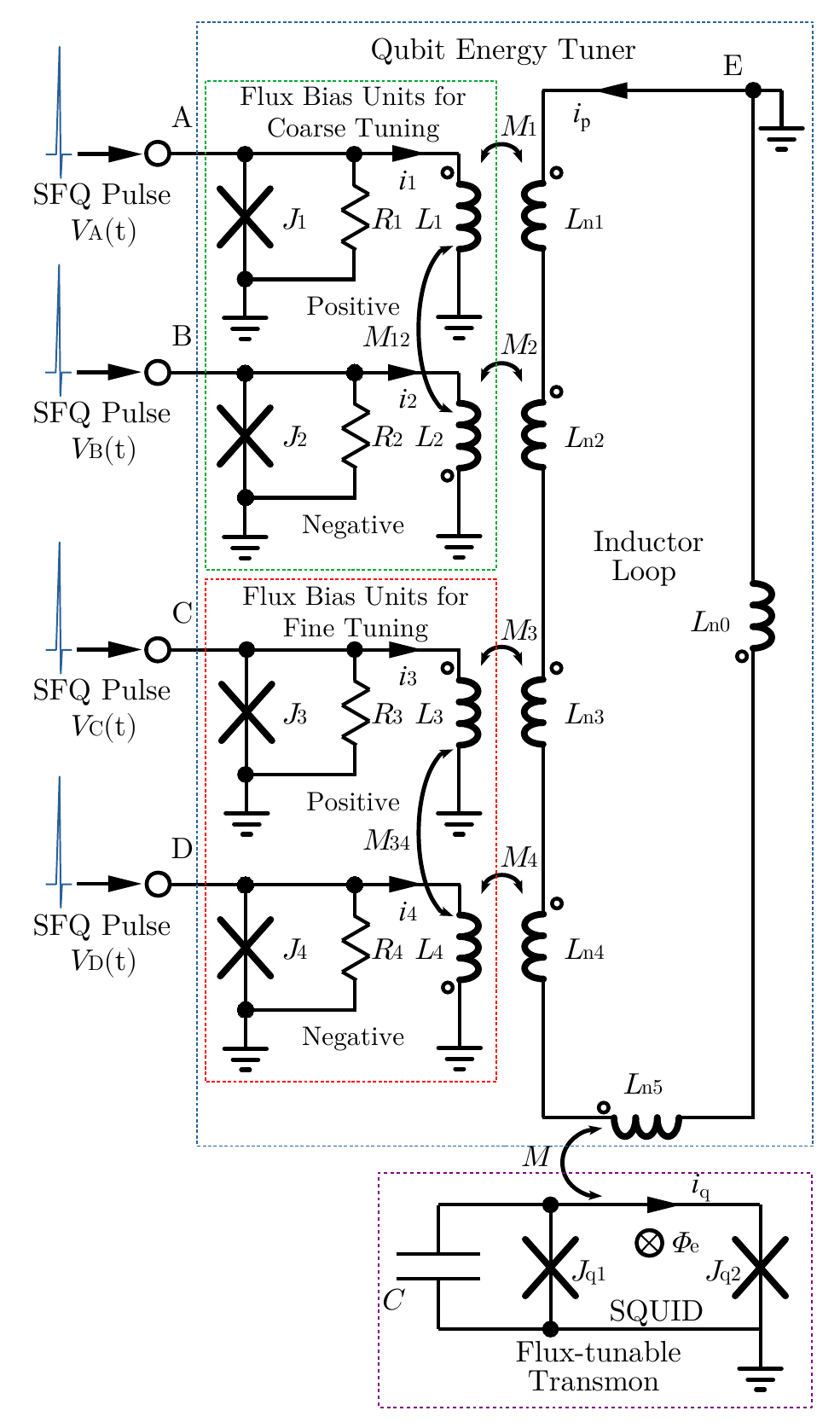}
\caption{\label{qet} Schematic of a qubit energy tuner with two pairs of flux bias units for coarse tuning and fine tuning.}
\end{center}
\end{figure}

\begin{figure}
\begin{center}
\includegraphics[scale=0.5]{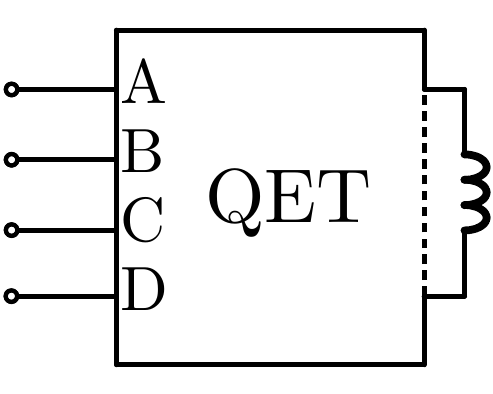}
\caption{\label{qet_symbol} Symbol of a QET.}
\end{center}
\end{figure}

\par
The QET shown in FIG.~\ref{qet} is taken as an example for the following analysis. It has a pair of flux bias units for coarse tuning and another pair for fine tuning. The parameters of elements in the example are listed in Table~\ref{tab:QETpara}. The symbol for QET in FIG.~\ref{qet} is drawn as FIG.~\ref{qet_symbol}. The reason why this kind of QET with two pairs of flux bias units is chosen to be analyzed is that it combines the accuracy, simplicity and speed better than other cases with only one or over two pairs of flux bias units. On the one hand, the QET with only a pair of flux bias units has only one precision, which causes a low speed of high-precision tuning or a low precision of high-speed tuning. On the other hand, the QET with three or more pairs of flux bias units have more ports and circuit elements, which means more complicated control, reduced reliability and larger footprint.
\par
According to the formula derivation in Appendix A, by ignoring the influence of SQUID, the current of the inductor loop $i_{\rm p}(t)$ is approximately equal to
\begin{equation}
\begin{split}
i_{\rm p}(t)=&\dfrac{1}{F}[(\varPhi_{\rm A}-\varPhi_{\rm B})(L_{\rm f}^{2}-M_{34}^{2})(L_{\rm c}+M_{12})M_{\rm c}\\
&+(\varPhi_{\rm C}-\varPhi_{\rm D})(L_{\rm c}^{2}-M_{12}^{2})(L_{\rm f}+M_{34})M_{\rm f}], \label{ipt}
\end{split}
\end{equation}
\noindent
where $\varPhi_{\rm A}$, $\varPhi_{\rm B}$, $\varPhi_{\rm C}$ and $\varPhi_{\rm D}$ are the integral of the voltage at node A, B, C and D over time $t$ respectively. 
Because the input signal to these nodes is SFQ, $\varPhi_{\rm A}$, $\varPhi_{\rm B}$, $\varPhi_{\rm C}$ and $\varPhi_{\rm D}$ are multiples of flux quantum $\Phi_{0}$.
$F$, $M_{\rm c}$, $M_{\rm f}$, $L_{\rm c}$ and $L_{\rm f}$ are defined in Appendix A.
\par
Then, the relationship between $i_{\rm p}(t)$ and the external flux through the SQUID $\varPhi_{\rm e}$ is 
\begin{equation}
\varPhi_{\rm e} = Mi_{\rm p}(t).
\end{equation}
Denoting
\begin{equation}
\varPhi_{\rm A}-\varPhi_{\rm B}=n_{\rm c}\Phi_{0}, \label{PhiAPhiBncPhi0}
\end{equation}
\begin{equation}
\varPhi_{\rm C}-\varPhi_{\rm D}=n_{\rm f}\Phi_{0},
\end{equation}
\begin{equation}
\dfrac{M}{F}(L^{2}_{\rm f}-M^{2}_{34})(L_{\rm c}+M_{12})M_{\rm c}=r_{\rm c}, \label{dfracMFEqrc}
\end{equation}
\begin{equation}
\dfrac{M}{F}(L^{2}_{\rm c}-M^{2}_{12})(L_{\rm f}+M_{34})M_{\rm f}=r_{\rm f}, \label{dfracMFEqrf}
\end{equation}
\begin{equation}
\dfrac{1}{F}(L^{2}_{\rm f}-M^{2}_{34})(L_{\rm c}+M_{12})M_{\rm c}\Phi_{0}=\Delta i_{\rm pc}, \label{deltaipc}
\end{equation}
\begin{equation}
\dfrac{1}{F}(L^{2}_{\rm c}-M^{2}_{12})(L_{\rm f}+M_{34})M_{\rm f}\Phi_{0}=\Delta i_{\rm pf}, \label{deltaipf}
\end{equation}
\begin{equation}
M\Delta i_{\rm pc}=\varPhi_{\rm ec}, \label{MipcPhiec}
\end{equation}
\begin{equation}
M\Delta i_{\rm pf}=\varPhi_{\rm ef}, \label{MipfPhief}
\end{equation}

\noindent
yields

\begin{equation}
i_{\rm p} = n_{\rm c}\Delta{i_{\rm pc}}+n_{\rm f}\Delta{i_{\rm pf}}, \label{ipEqncDeltaipcPlusnfDeltapf}
\end{equation}

\begin{equation}
\varPhi_{\rm e} = n_{\rm c}\varPhi_{\rm ec}+n_{\rm f}\varPhi_{\rm ef}, \label{PhieEqncPhiecnfPhief}
\end{equation}

\begin{equation}
\varPhi_{\rm ec} = r_{\rm c}\Phi_{0}, \label{PhiecrcPhi0}
\end{equation}

\begin{equation}
\varPhi_{\rm ef} = r_{\rm f}\Phi_{0}. \label{PhiefrcPhi0}
\end{equation}

\noindent
Equation (\ref{PhieEqncPhiecnfPhief}),(\ref{PhiecrcPhi0}) and (\ref{PhiefrcPhi0}) mean that the flux provided by QET can be divided in to two parts, $n_{\rm c}\varPhi_{\rm ec}$ and $n_{\rm f}\varPhi_{\rm ef}$, which are respectively created by coarse tuning and fine tuning. $\varPhi_{\rm ec}$ can be regarded as the flux unit of coarse tuning and $\varPhi_{\rm ef}$ can be regarded as the flux unit of fine tuning. If $n_{\rm c}$ (or $n_{\rm f}$) SFQ pulses are inputted to port A (or C) of the QET, then the external flux through the SQUID will increase by $n_{\rm c}$ times of $\varPhi_{\rm ec}$ (or $n_{\rm f}$ times of $\varPhi_{\rm ef}$). Then, if this external flux needs to be eliminated, $n_{\rm c}$ (or $n_{\rm f}$) SFQ pulses should be inputted to port B (or D). Usually for fine tuning $r_{\rm f}$ is smaller than $r_{\rm c}$. If
\begin{equation}
L_{\rm c} = L_{1}= L_{\rm n1}=L_{2}=L_{\rm n2}, \label{L_cEq}
\end{equation}
\begin{equation}
L_{\rm f} = L_{3}= L_{\rm n3}=L_{4}=L_{\rm n4}, \label{L_fEq}
\end{equation}
then the ratio of the flux unit of coarse tuning to the flux unit of fine tuning can be defined as
\begin{equation}
r_{\rm cf} = \dfrac{\varPhi_{\rm ec}}{\varPhi_{\rm ef}}. \label{rcfdefine}
\end{equation}
\noindent
With Equation (\ref{PhiAPhiBncPhi0})$\sim$(\ref{rcfdefine}), there is
\begin{equation}
r_{\rm cf} = \dfrac{r_{\rm c}}{r_{\rm f}} = \dfrac{\Delta i_{\rm pc}}{\Delta i_{\rm pf}} = \dfrac{K_{\rm c}(1-K_{34})}{K_{\rm f}(1-K_{12})}, \label{rcf}
\end{equation}
\noindent
where the coupling coefficients are
\begin{equation}
K_{\rm c} = \dfrac{M_{\rm c}}{\sqrt{L_{1}L_{\rm n1}}} = \dfrac{M_{\rm c}}{\sqrt{L_{2}L_{\rm n2}}} =  \dfrac{M_{\rm c}}{L_{\rm c}}, 
\end{equation}
\begin{equation}
K_{\rm f} = \dfrac{M_{\rm f}}{\sqrt{L_{1}L_{\rm n1}}} = \dfrac{M_{\rm f}}{\sqrt{L_{2}L_{\rm n2}}} =  \dfrac{M_{\rm f}}{L_{\rm f}},
\end{equation}
\begin{equation}
K_{\rm 12} = \dfrac{M_{\rm 12}}{\sqrt{L_{1}L_{\rm 2}}} = \dfrac{M_{\rm 12}}{L_{\rm c}},
\end{equation}
\begin{equation}
K_{\rm 34} = \dfrac{M_{\rm 34}}{\sqrt{L_{3}L_{\rm 4}}} = \dfrac{M_{\rm 34}}{L_{\rm f}}. \label{K34}
\end{equation}
\par
The parameter $r_{\rm cf}$ means the ratio of the flux precision of coarse tuning to that of fine tuning. It should have an appropriate value larger than $1$, like 10, to distinguish the two precisions. The parameter $r_{\rm f}$ is the ratio of the smallest variation of the flux $\varPhi_{\rm e}$, $\varPhi_{\rm ef}$, to flux quantum $\Phi_0$, that is, it determines the flux precision of fine tuning. The parameter $r_{\rm c}$ is the ratio of $\varPhi_{\rm ec}$ to flux quantum $\Phi_0$ and can be set by $r_{\rm c}=r_{\rm cf}\cdot r_{\rm f}$. 
\par
To design a QET, the parameters $r_{\rm cf}$, $r_{\rm f}$ and $r_{\rm c}$ are main concerns and should be firstly determined. Then, with constraints including Equation (\ref{dfracMFEqrc}), (\ref{dfracMFEqrf}), (\ref{L_cEq}), (\ref{L_fEq}), (\ref{rcf})$\sim$(\ref{K34}), all parameter values of circuit elements should be tried and iterated to meet the requirements from the higher-level design, for example, the footprint of the QET on the chip is matched with the footprint of the qubit.

\section{Ideal Z control by Square-wave-like Currents}

For a flux-tunable transmon, the ideal case for Z control is that the waveforms of currents producing external flux $\varPhi_{\rm e}$ are square-wave-like. In this section how the ideal Z control is performed is discussed.
\par
The Hamiltonian of a flux-tunable transmon \cite{koch2007charge} is
\begin{equation}
    \hat{H}=4E_{\rm C}\left(\hat{n}-n_{\rm g}\right)^{2}-E_{\rm JS}(\varphi_{\rm e}){\rm cos}(\hat{\phi}), \label{transmonH}
\end{equation}
\noindent
where
\begin{equation}
    E_{\rm JS}(\varphi_{\rm e})=E_{\rm J\Sigma}|\cos(\varphi_{\rm e})|\sqrt{1+d^{2} \tan^{2}(\varphi_{\rm e})},
\end{equation}
\noindent
is the effective Josephson energy of the SQUID of the transmon with a total Josephson coupling energy of two junctions
\begin{equation}
    E_{\rm J\Sigma}=E_{\rm J1}+E_{\rm J2},
\end{equation}
\noindent
a asymmetry coefficient
\begin{equation}
    d=\dfrac{E_{\rm J2}-E_{\rm J1}}{E_{\rm J\Sigma}},
\end{equation}
and a reduced external flux
\begin{equation}
    \varphi_{\rm e} = \pi \dfrac{\varPhi_{\rm e}(t)}{\Phi_{0}}.
\end{equation}
\noindent
$E_{\rm C}$ is the charging energy of the transmon. $n_{\rm g}$ is the effective offset charge. $\hat{n}$ and $\hat{\phi}$ are respectively the number operator and the phase operator of Cooper pairs. For convenience, the hats of all operators including Hamiltonian are left out in the following derivation. 
\par
The solution for the $k^{\rm th}$ eigen energy of Equation (\ref{transmonH}) with first order approximation of perturbation theory \cite{koch2007charge} is
\begin{equation}
    E_{k}=k\sqrt{8E_{\rm C}E_{\rm JS}(\varphi_{\rm e})}-\dfrac{E_{\rm C}}{12}(6k^2+6k+3)-E_{\rm JS}(\varphi_{\rm e}). \label{Ekdefine}
\end{equation}
\noindent
Usually, the external flux $\varPhi_{\rm e}(t)$ at the moment $t$ for Z control is provided by a conductor line besides the SQUID with its current, $i_{\rm z}(t)$, and a mutual inductance between the line and the SQUID, $M$. Because the current in the SQUID is much smaller than $i_{\rm z}(t)$, its influence on $i_{\rm z}(t)$ can be ignored and there is
\begin{equation}
    \varPhi_{\rm e}(t)=Mi_{\rm z}(t).
\end{equation}
\noindent
Therefore, $E_{\rm JS}(\varphi_{\rm e})$ can be written as
\begin{equation}
    E_{\rm JS}(i_{\rm z}(t))=E_{\rm J\Sigma}\left|\cos(\pi \dfrac{Mi_{\rm z}(t)}{\Phi_{0}})\right|\sqrt{1+d^{2} \tan^{2}(\pi \dfrac{Mi_{\rm z}(t)}{\Phi_{0}})}. \label{EJSDefine}
\end{equation}
\noindent
By changing $i_{\rm z}(t)$, $E_{\rm JS}(i_{\rm z}(t))$ can be set to a target value, then the energy level $E_{\rm k}$, especially $E_0$ and $E_1$ can be tuned so that qubit frequency is set to the corresponding target value. When $E_{\rm JS}(i_{\rm z}(t))$ is set, the condition $E_{\rm JS}(i_{\rm z}(t))/E_{\rm C}>>1$ should be guaranteed to make sure that the qubit is transmon. According to Equation (\ref{Ekdefine}), we have
\begin{equation}
    E_0=-\dfrac{E_{\rm C}}{4}-E_{\rm JS}(\varphi_{\rm e}),
\end{equation}
\begin{equation}
    E_1=\sqrt{8E_{\rm C}E_{\rm JS}(\varphi_{\rm e})}-E_{\rm C}-\dfrac{E_{\rm C}}{4}-E_{\rm JS}(\varphi_{\rm e}),
\end{equation}
\begin{equation}
    E_2=2\sqrt{8E_{\rm C}E_{\rm JS}(\varphi_{\rm e})}-3E_{\rm C}-\dfrac{E_{\rm C}}{4}-E_{\rm JS}(\varphi_{\rm e}).
\end{equation}
\noindent
Therefore, the differences of energy levels are
\begin{equation}
\begin{split}
E_{10} = E_1-E_0 = \sqrt{8E_{\rm C}E_{\rm JS}(\varphi_{\rm e})}-\ \ E_{\rm C},
\end{split}
\end{equation}
\begin{equation}
\begin{split}
E_{21} = E_2-E_1 = \sqrt{8E_{\rm C}E_{\rm JS}(\varphi_{\rm e})}-2E_{\rm C},
\end{split}
\end{equation}
and the anharmonicity of the qubit is
\begin{equation}
\begin{split}
\alpha = E_{21}-E_{10} = -E_{\rm C}.
\end{split}
\end{equation}
Without losing generality, $i_{\rm z}(t)$ has a square-wave-like waveform and is set to be
\begin{equation}
    i_{\rm z}(t)=
    \begin{cases}
    i_{\rm w}, &t_{\rm s} \leqslant t \leqslant t_{\rm e}, \\
    i_{\rm i}, &0 \leqslant t < t_{\rm s} \ {\rm or}\  t \textgreater t_{\rm e}, \\
    \end{cases}
\end{equation}
\noindent
where $i_{\rm w}$ and $i_{\rm i}$ are the currents for setting working frequency $\omega_{\rm qw}$ and idle frequency $\omega_{\rm qi}$ of a transmon, respectively. $\omega_{\rm qw}$ is the qubit frequency used for Z control. $\omega_{\rm qi}$ is the qubit frequency when it is idle and is determined to be the frequency of the rotating frame \cite{krantz2019quantum}. $t_{\rm s}$ and $t_{\rm e}$ are the moments when a gate operation starts and ends, respectively. Then, the qubit frequency turns to be
\begin{equation}
\omega_{\rm q}(t)=
\begin{cases}
\omega_{\rm qw}, &t_{\rm s} \leqslant t \leqslant t_{\rm e}, \\
\omega_{\rm qi}, &0 \leqslant t < t_{\rm s} \ {\rm or}\  t \textgreater t_{\rm e}, \\
\end{cases}
\end{equation}
where
\begin{equation}
\omega_{\rm qw}=\left(\sqrt{8E_{\rm C}E_{\rm JS}(i_{\rm w})}-E_{\rm C}\right)/\hbar, \label{omegaqwEq}
\end{equation}
\begin{equation}
\omega_{\rm qi}=\left(\sqrt{8E_{\rm C}E_{\rm JS}(i_{\rm i})}-E_{\rm C}\right)/\hbar. \label{omegaqiEq}
\end{equation}
\noindent
Here, we denote
\begin{equation}
\Delta\omega_{\rm q}=\omega_{\rm qw}-\omega_{\rm qi}. \label{Deltaomegaqomegaqwmomegaqi}
\end{equation}
\noindent
With Equation (\ref{omegaqwEq}), (\ref{omegaqiEq}) and (\ref{Deltaomegaqomegaqwmomegaqi}), we have
\begin{equation}
\Delta\omega_{\rm q}=\sqrt{8E_{\rm C}}\left(\sqrt{E_{\rm JS}(i_{\rm w})}-\sqrt{E_{\rm JS}(i_{\rm i})}\right). \label{DeltaomegaqEqsqrt8EC}
\end{equation}
\par
For an idle qubit, its Hamiltonian is
\begin{equation}
H_0 = \hbar\left(\omega_{\rm qi}a^{\dagger}a+\dfrac{\alpha}{2}a^{\dagger}a^{\dagger}aa\right).
\end{equation}
\noindent
Actually, the time-dependent Hamiltonian of the qubit is
\begin{equation}
H = \hbar\left(\Delta\omega\left(t\right) a^{\dagger}a+\omega_{\rm qi}a^{\dagger}a+\dfrac{\alpha}{2}a^{\dagger}a^{\dagger}aa\right),
\end{equation}
\noindent
where $a^{\dagger}$ and $a$ are the creation operator and the annihilation operator, respectively, and $\Delta\omega(t)$ is defined by
\begin{equation}
    \Delta\omega(t) = \omega_{\rm q}(t)-\omega_{\rm qi}.
\end{equation}
$\omega_{\rm q}(t)$ is the actual frequency of the qubit. 
We denote
\begin{equation}
H_{\rm dz} = \hbar\Delta\omega\left(t\right) a^{\dagger}a, \label{HdzEqhbarDeltaomegatadaga}
\end{equation}
\noindent as the drive Hamiltonian for Z control, so there is
\begin{equation}
H = H_0 + H_{\rm dz}. \label{HEqH0plusHdz}
\end{equation}
\noindent
In the rotating frame, the drive Hamiltonian for Z control turns to be
\begin{equation}
\widetilde{H} =  H_{\rm dz}, \label{widetildeHEqHdz}
\end{equation}
\noindent
and the corresponding evolution operator in the rotating frame is
\begin{equation}
\widetilde{U}_{\rm dz} = \mathcal{T}{\rm exp}\left(-\mathrm{i}\int_{t_{\rm s}}^{t_{\rm e}}\dfrac{\widetilde{H}}{\hbar}{\rm d}t\right), \label{Udzdefine}
\end{equation}
where $\mathcal{T}$ is chronological operator. With Equation (\ref{HdzEqhbarDeltaomegatadaga}), (\ref{widetildeHEqHdz}) and (\ref{Udzdefine}), we have
\begin{equation}
\widetilde{U}_{\rm dz} = \mathcal{T}{\rm exp}\left(-\mathrm{i}a^\dagger a\int_{t_{\rm s}}^{t_{\rm e}}\Delta\omega\left(t\right){\rm d}t\right). \label{Udzapa}
\end{equation}
\noindent
In the ideal situation where $i_{\rm w}$ and $i_{\rm i}$ are constants, when the qubit is working ($t_{\rm s} \leqslant t \leqslant t_{\rm e}$), there is $\omega_{\rm q}(t)=\omega_{\rm qw}$, so $\Delta\omega(t)$ becomes the constant $\Delta\omega_{\rm q}$:
\begin{equation}
\Delta\omega(t)=\Delta\omega_{\rm q}=\omega_{\rm qw}-\omega_{\rm qi}, \label{DeltaomegatDeltaomegaqomegaqwomegaqi}
\end{equation}
and the Hamiltonian $H$ turns to be
\begin{equation}
H = H_{\rm w} = \hbar\left(\omega_{\rm qw}a^{\dagger}a+\dfrac{\alpha}{2}a^{\dagger}a^{\dagger}aa\right).
\end{equation}
\noindent
We define
\begin{equation}
\varphi = -\int_{t_{\rm s}}^{t_{\rm e}}\Delta\omega\left(t\right){\rm d}t \label{definephaseshift}
\end{equation}
\noindent
as the phase shift realized by Z control, and define
\begin{equation}
t_{\rm z} = t_{\rm e} - t_{\rm s} \label{definetZ}
\end{equation}
\noindent
as the gate operation time for Z control.
With Equation (\ref{DeltaomegaqEqsqrt8EC}), (\ref{DeltaomegatDeltaomegaqomegaqwomegaqi}), (\ref{definephaseshift}) and (\ref{definetZ}), we have
\begin{equation}
\begin{split}
\varphi =& -\Delta\omega_{\rm q}t_{\rm z}\\
        =& \sqrt{8E_{\rm C}}\left(\sqrt{E_{\rm JS}(i_{\rm i})}-\sqrt{E_{\rm JS}(i_{\rm w})}\right)t_{\rm z}. \label{phiDeltaomegaqtz}
\end{split}
\end{equation}
\noindent
And according to Equation (\ref{Udzapa}), the corresponding evolution operator for the qubit turns to be
\begin{equation}
\begin{split}
\widetilde{U}_{\rm dz} = &
\left(
    \begin{matrix}
        1 & 0 \\
        0 & e^{\mathrm{i}\varphi}
    \end{matrix}
\right)\\
=&
\left(
    \begin{matrix}
        1 & 0 \\
        0 & \exp\left(\mathrm{i}\sqrt{8E_{\rm C}}\left(\sqrt{E_{\rm JS}(i_{\rm i})}-\sqrt{E_{\rm JS}(i_{\rm w})}\right)t_{\rm z}\right)
    \end{matrix}
\right).
\end{split}
\end{equation}
To realize on-chip Z control by SFQ, instead of chosing Z control line, $i_{\rm i}$ and $i_{\rm w}$ can be produced by the inductor loop current $i_{\rm p}(t)$ of a QET, which means making $i_{\rm z}(t)=i_{\rm p}(t)$.

\section{Simulation about a QET and Its Gate Operations}

\subsection*{A. a Single QET}

\begin{figure*}[htb]
\begin{center}
\includegraphics[scale=0.6]{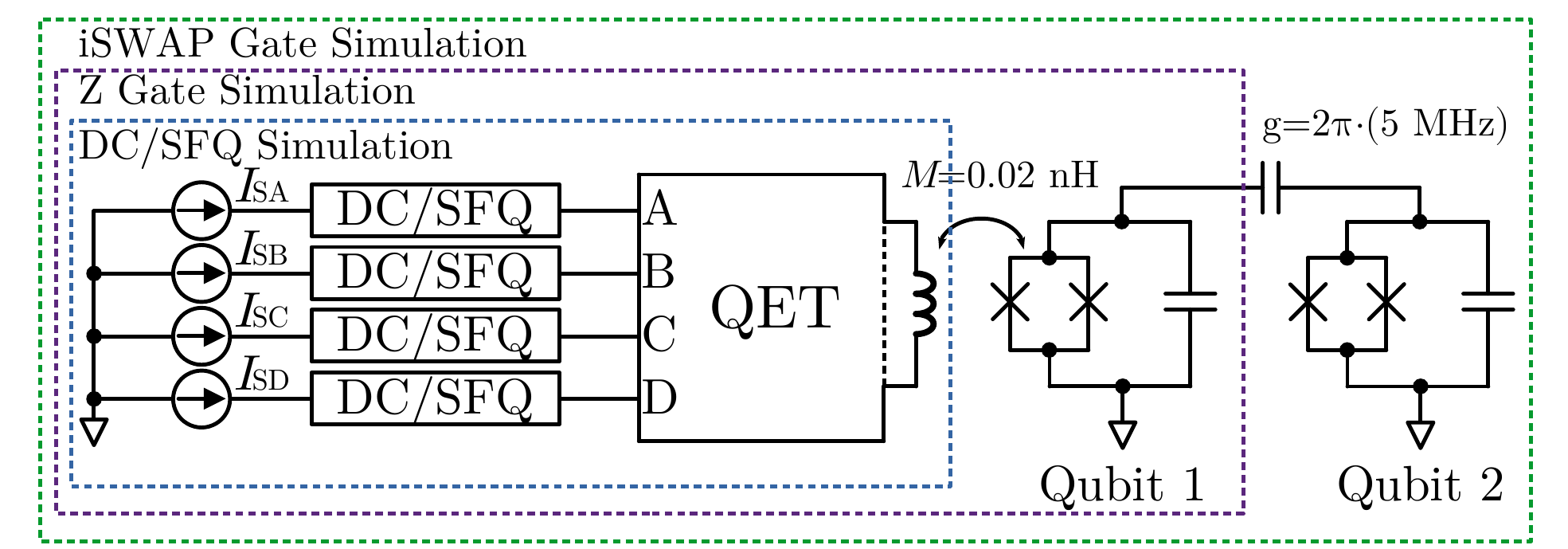}
\caption{\label{qet_dcsfq_simcircuit} The circuits for all simulations.}
\end{center}
\end{figure*}

\begin{figure}[htb]
\begin{center}
\includegraphics[scale=0.5]{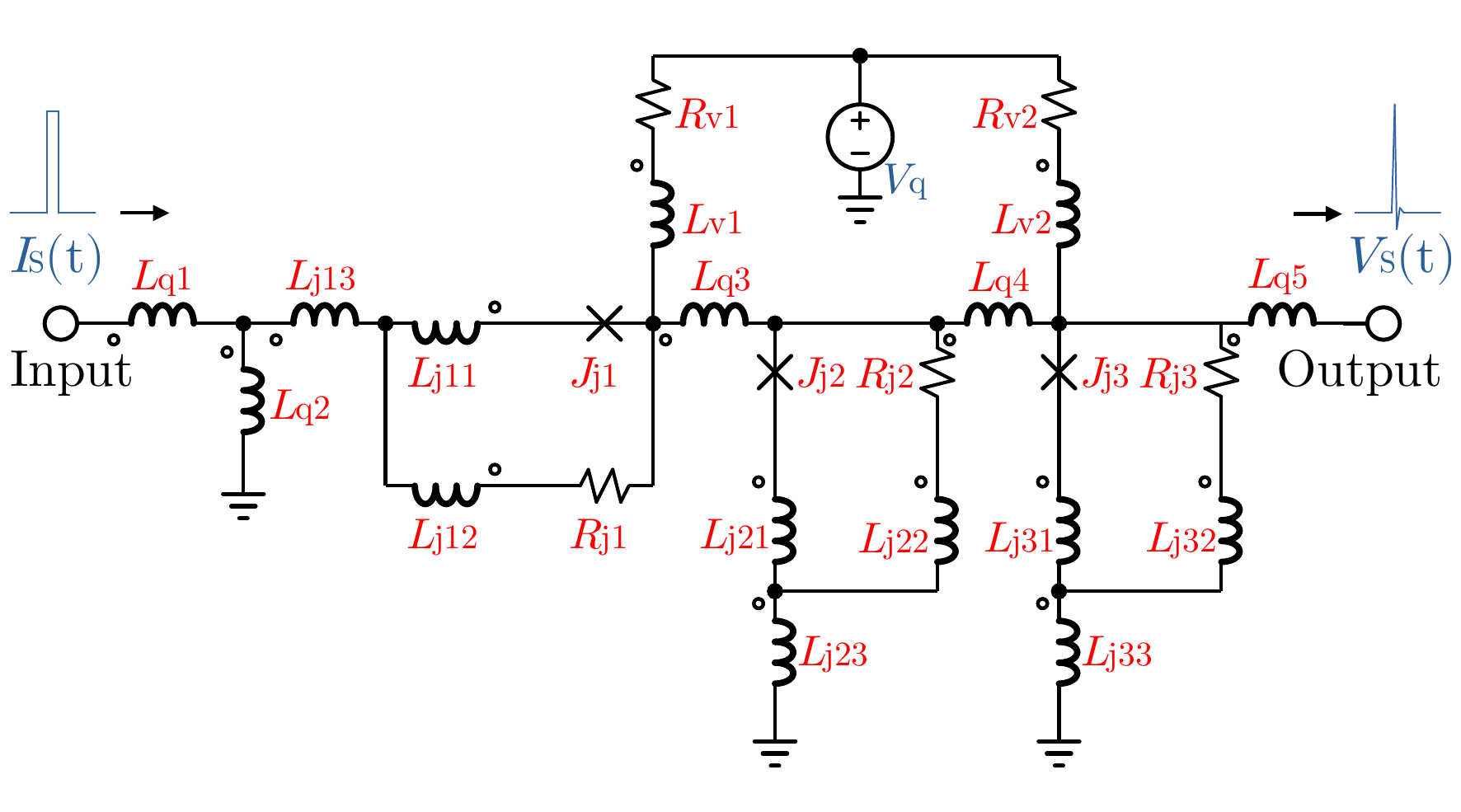}
\caption{\label{dcsfq_simcircuit} The simulation circuits of DC/SFQ.}
\end{center}
\end{figure}

\begin{figure}
\begin{center}
\includegraphics[scale=0.9]{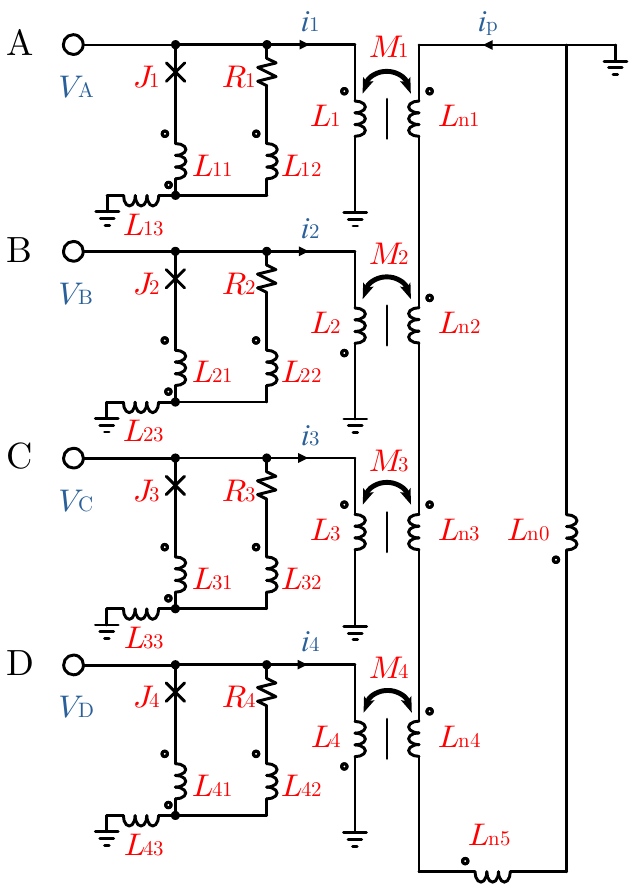}
\caption{\label{qet_simcircuit} The simulation circuits of QET.}
\end{center}
\end{figure}

\begin{figure*}[htb]
\begin{center}
\includegraphics[scale=0.65]{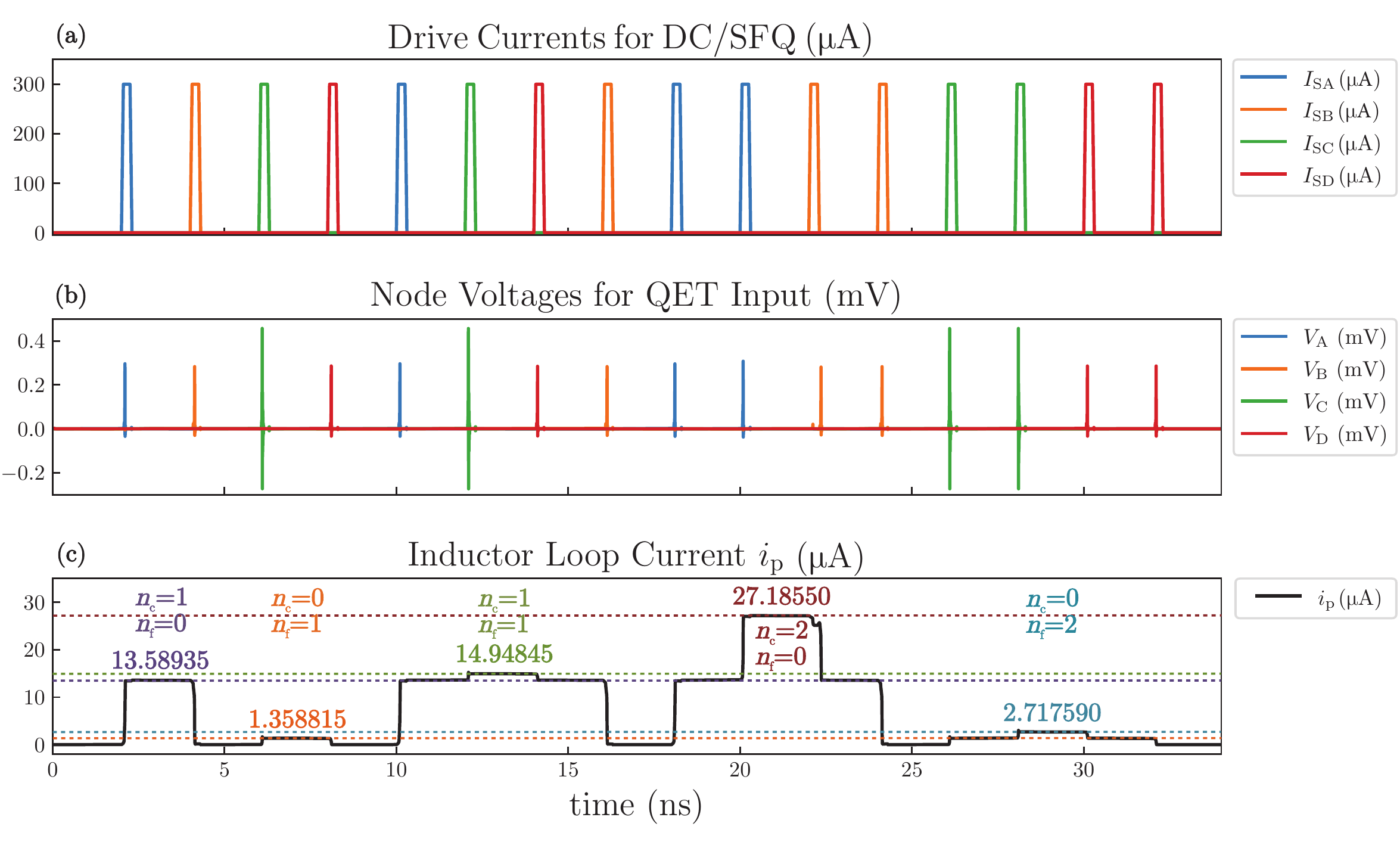}
\caption{\label{qet_simulation_result} The responses of QET to SFQ pulses produced by DC/SFQ in simulation.}
\end{center}
\end{figure*}

\begin{table}
\caption{The parameters of elements in the DC/SFQ.}
\label{tab:DCSFQpara}
\begin{ruledtabular}
\begin{tabular}{cccc}
Parameters      & Values    & Parameters        & Values \\
\cline{1-4}
$L_{\rm q1}$	& 1.071 pH  & $L_{\rm j33}$     & 0.103 pH \\
$L_{\rm q2}$    & 3.927 pH  & $J_{\rm j1}$      & 225 $\mu$A \\
$L_{\rm q3}$    & 0.913 pH  & $J_{\rm j2}$      & 225 $\mu$A \\
$L_{\rm q4}$    & 4.399 pH  & $J_{\rm j3}$      & 250 $\mu$A \\
$L_{\rm q5}$    & 1.090 pH  & $L_{\rm v1}$      & 16.8 pH \\
$L_{\rm j11}$   & 0.058 pH  & $L_{\rm v2}$      & 15.5 pH \\
$L_{\rm j12}$   & 0.945 pH  & $R_{\rm v1}$      & 9.09 $\Omega$ \\
$L_{\rm j13}$   & 0.355 pH  & $R_{\rm v2}$      & 14.29 $\Omega$ \\
$L_{\rm j21}$   & 0.05 pH   & $R_{\rm j1}$      & 0.766 $\Omega$ \\
$L_{\rm j22}$   & 0.955 pH  & $R_{\rm j2}$      & 0.766 $\Omega$ \\
$L_{\rm j23}$   & 0.096 pH  & $R_{\rm j3}$      & 0.688 $\Omega$ \\
$L_{\rm j31}$   & 0.028 pH  & $V_{\rm q}$       & 2.5 mV \\
$L_{\rm j32}$   & 0.961 pH  & \                 & \  \\
\end{tabular}
\end{ruledtabular}
\end{table}

\begin{table}
\caption{The parameters of elements in the QET.}
\label{tab:QETpara}
\begin{ruledtabular}
\begin{tabular}{cccc}
Parameters      & Values        & Parameters    & Values \\
\cline{1-4}
$L_{1}$         & $L_{\rm c}$   & $L_{\rm 11}$  & 0.05 pH \\
$L_{2}$         & $L_{\rm c}$   & $L_{\rm 12}$  & 0.955 pH \\
$L_{3}$         & $L_{\rm f}$   & $L_{\rm 13}$  & 0.096 pH \\
$L_{4}$         & $L_{\rm f}$   & $L_{\rm 21}$  & 0.05 pH \\
$L_{\rm c}$     & 10 nH         & $L_{\rm 22}$  & 0.955 pH \\
$L_{\rm f}$     & 10 nH         & $L_{\rm 23}$  & 0.096 pH \\
$L_{\rm n0}$    & 1  nH         & $L_{\rm 31}$  & 0.05 pH \\
$L_{\rm n1}$    & $L_{\rm c}$   & $L_{\rm 32}$  & 0.955 pH \\
$L_{\rm n2}$    & $L_{\rm c}$   & $L_{\rm 33}$  & 0.096 pH \\
$L_{\rm n3}$    & $L_{\rm f}$   & $L_{\rm 41}$  & 0.05 pH \\
$L_{\rm n4}$    & $L_{\rm f}$   & $L_{\rm 42}$  & 0.955 pH \\
$L_{\rm n5}$    & 2 nH          & $L_{\rm 43}$  & 0.096 pH \\
$M_{1}$         & $M_{\rm c}$   & $R_{1}$       & 0.766 $\Omega$ \\
$M_{2}$         & $M_{\rm c}$   & $R_{2}$       & 0.766 $\Omega$ \\
$M_{3}$         & $M_{\rm f}$   & $R_{3}$       & 0.766 $\Omega$ \\
$M_{4}$         & $M_{\rm f}$   & $J_{1}$       & 160 $\mu$A \\
$M_{12}$        & 7.023 nH      & $J_{2}$       & 160 $\mu$A \\
$M_{34}$        & 7.023 nH      & $J_{3}$       & 160 $\mu$A \\
$M_{\rm c}$     & 8 nH          & $J_{4}$       & 160 $\mu$A \\
$M_{\rm f}$     & 0.8 nH        & $M$           & 0.02 nH \\
\end{tabular}
\end{ruledtabular}
\end{table}

\par
In order to show how the inductor loop current $i_{\rm p}(t)$ of a QET is controlled by SFQ signal, a simulation with supreconducting circuit simulation software WRSpice for the circuits in the blue-dashed-line box in FIG.~\ref{qet_dcsfq_simcircuit} is performed. The SFQ pulses sent to the input ports of QET, A, B, C and D, are generated by four DC/SFQs, each of which is drived by a time-dependent current source. Here, the DC/SFQ is only used for generating SFQ pulses to verify the functionality of the QET. In practical engineering, the QET can also be drived by SFQ pulses from other SFQ digital circuits. The existance and influence of two qubits in FIG.~\ref{qet_dcsfq_simcircuit} are ignored temporarily. The circuits of DC/SFQ and QET for simulation is shown in FIG.~\ref{dcsfq_simcircuit} and FIG.~\ref{qet_simcircuit}, and the corresponding parameters of their elements are listed in TABLE~\ref{tab:DCSFQpara} and TABLE~\ref{tab:QETpara}, which are basd on the SFQ circuit design data from Ref.~\onlinecite{ligang2018phd}. In these two tables, the parameters whose names start with letter ``$J$'' are the critical currents of corresponding Josephson junctions. The circuit of the QET in FIG.~\ref{qet_simcircuit} is a little different from FIG.~\ref{qet} for consideration of the parasitic inductances, but the functions of the QET will not change essentially. FIG.~\ref{qet_simulation_result} shows the simulation results including the waveforms of (a) the drive currents for DC/SFQ, $I_{\rm SA}$, $I_{\rm SB}$, $I_{\rm SC}$ and $I_{\rm SD}$ mentioned in FIG.~\ref{qet_simcircuit}; (b) the node voltages for QET input ports, $V_{\rm A}$, $V_{\rm B}$, $V_{\rm C}$ and $V_{\rm D}$, which are SFQ pulses with time interval 2 ns; (c) the inductor loop current $i_{\rm p}$. 

\begin{table}
    \caption{The simulation values and analytical values of the inductor loop currents with different $n_{\rm c}$ and $n_{\rm f}$ in the simulation for a single QET.}
    \label{tab:inductorloopcurrent_qet}
    \begin{ruledtabular}
    \begin{tabular}{ccccc}
    $n_{\rm c}$     & $n_{\rm f}$   & Simulation ($\mu$A)    & Analytical ($\mu$A) & Relative Deviation (\%)\\
    \cline{1-5}
    1               & 0             & 13.58935           & 13.03599        & 4.245         \\
    0               & 1             & 1.358815           & 1.303599        & 4.236         \\
    1               & 1             & 14.94845           & 14.33959        & 4.246         \\
    2               & 0             & 27.18550           & 26.07198        & 4.271         \\
    0               & 2             & 2.717590           & 2.607198        & 4.234         \\
    \end{tabular}
    \end{ruledtabular}
\end{table}
\par
To change the inductor loop current $i_{\rm p}(t)$ which provides external flux $\varPhi_{\rm e}=Mi_{\rm p}(t)$, $n_{\rm c}$ and $n_{\rm f}$ can be set by the SFQ pulse sequence from DC/SFQ. The simulation and analytical values of the inductor loop current with different $n_{\rm c}$ and $n_{\rm f}$ are compared in TABLE~\ref{tab:inductorloopcurrent_qet}. First, by setting $n_{\rm c}=1$ and $n_{\rm f}=0$ with SFQ pulses sent to port A and port B (coarse tuning), $\Delta i_{\rm pc}$ can be extracted from the height of the leftmost lug boss of the inductor loop current curve in FIG.~\ref{qet_simulation_result}(c). The extraction value of $\Delta i_{\rm pc}$ is 13.58935 ${\rm \mu A}$, which is close to the value 13.03599 ${\rm \mu A}$ calculated by Equation (\ref{deltaipc}) with relative deviation 4.245\%. Similarly, by setting $n_{\rm c}=0$ and $n_{\rm f}=1$ with SFQ pulses sent to port C and port D (fine tuning), $\Delta i_{\rm pf}$ can also be extracted as 1.358815 ${\rm \mu A}$ from the second left current lug boss, which is also close to analytical solution 1.303599 ${\rm \mu A}$ from Equation (\ref{deltaipc}) with relative deviation 4.236\%. Therefore, the $r_{\rm cf}$ from this simulation is 10.00088 according to Equation (\ref{rcf}), almost the same as 10.0, the theory value from analytical solutions.
\par
By setting $n_{\rm c}=1$ and $n_{\rm f}=1$, the two parts of the inductor loop current correspondingly made by coarse tuning and fine tuning can be accumulated, as shown in the third left lug boss in FIG.~\ref{qet_simulation_result}(c). By setting $n_{\rm c}=2$ and $n_{\rm f}=0$, the inductor loop current can be double times of $\Delta i_{\rm pc}$. Similarly, by setting $n_{\rm c}=0$ and $n_{\rm f}=2$, the inductor loop current can also be double times of $\Delta i_{\rm pf}$. Generally, if the inductor loop current is required to be $N_{\rm c}$ times of $\Delta i_{\rm pc}$ plus $N_{\rm f}$ times of $\Delta i_{\rm pf}$, then $n_{\rm c}$ should be set as $N_{\rm c}$ and $n_{\rm f}$ should be set as $N_{\rm f}$ according to Equation~(\ref{ipEqncDeltaipcPlusnfDeltapf}).
\par
The waveform of the inductor loop current is similar with composited square waves on the whole, and their rising edges and falling edges are steep, which is helpful for avoiding the crosstalk when the qubit frequency is changing across frequencies of other qubits or resonators, because the qubit frequency is changed quickly enough within time (several picosecond) much shorter than a gate operation time (several nanosecond).

\subsection*{B. Z Gate by a QET}
\begin{figure*}[htb]
\begin{center}
\includegraphics[scale=0.7]{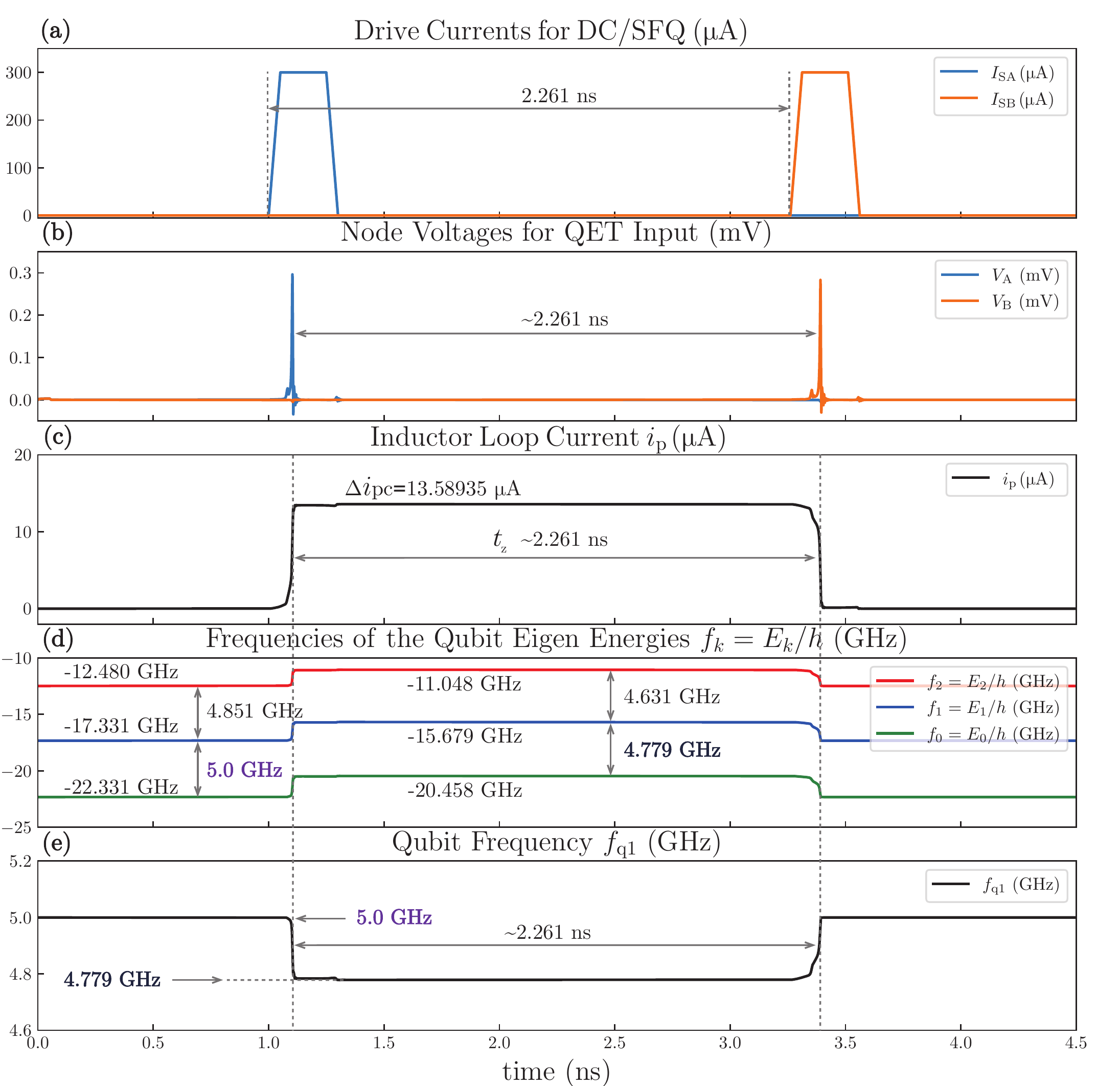}
\caption{\label{ZgateSimulationResult} Z Gate Simulation Results.}
\end{center}
\end{figure*}
\begin{figure}[htb]
\begin{center}
\includegraphics[scale=0.35]{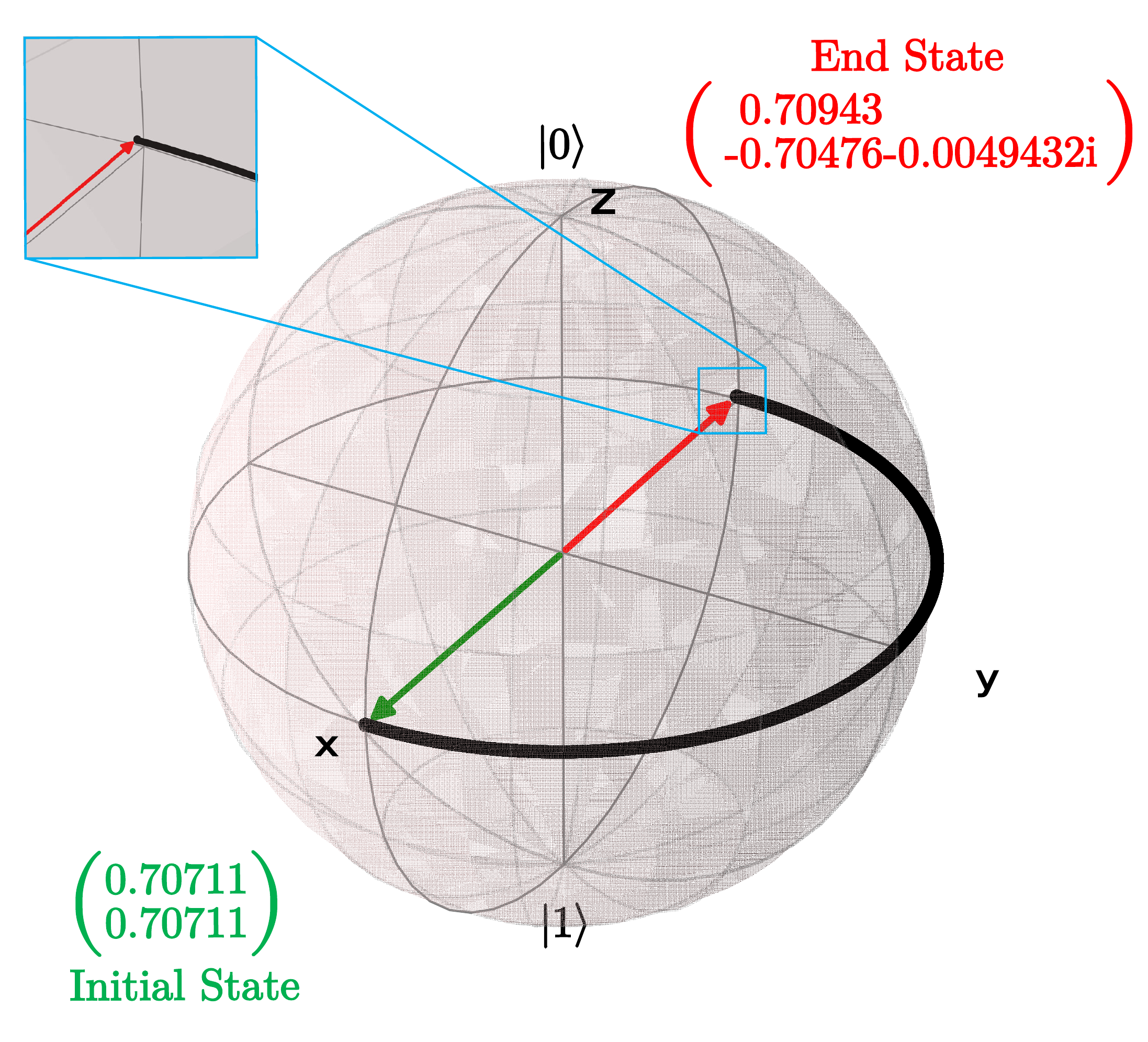}
\caption{\label{ZgateSimulationResultBlochSphere} Trajectory of the point representing qubit state in Z gate simulation (in black).}
\end{center}
\end{figure}
The simulation in this section shows how a QET can perform a Z gate. The circuit for simulation is defined as the circuit in the purple-dashed-line box of FIG.~\ref{qet_dcsfq_simcircuit}, which is based on the circuit of the former simulation in the blue-dashed-line box. The controlled qubit, Qubit 1, is a symmetric flux-tunable transmon connected to the former circuit, so we set $d=0$ and $E_{\rm J1}=E_{\rm J2}=E_{\rm J}$. Qubit 2 is ignored temporarily. By controlling the time interval of two SFQ pulses inputted to the port A and B of the QET, the phase of a flux-tunable transmon can be adjusted. Qubit 1 is only drived by coarse tuning with $n_{\rm c}=1$, so we set $i_{\rm i}=0$ and $i_{\rm w}=\Delta i_{\rm pc}$, and then $\Delta\omega(t)$ can be approximately treated as the constant $\Delta\omega_{\rm q}$ when $t_{\rm s} \leqslant t \leqslant t_{\rm e}$, that is,
\begin{equation}
    i_{\rm z}(t) = i_{\rm p}(t) =
    \begin{cases}
    \Delta i_{\rm pc}, &t_{\rm s} \leqslant t \leqslant t_{\rm e}, \\
    0, &0 \leqslant t < t_{\rm s} \ {\rm or}\  t \textgreater t_{\rm e}, \\
    \end{cases}
\end{equation}
and 
\begin{equation}
\Delta\omega_{\rm q} = \dfrac{4\sqrt{E_{\rm C}E_{\rm J}}}{\hbar}\left(\sqrt{\left|{\rm cos}\left(\pi\dfrac{M \Delta i_{\rm pc}}{\Phi_{0}}\right)\right|}-1\right). \label{Deltaomigaqexample}
\end{equation}
\noindent
Then, with Equation (\ref{phiDeltaomegaqtz}) and (\ref{Deltaomigaqexample}), we have
\begin{equation}
\varphi = \dfrac{4\sqrt{E_{\rm C}E_{\rm J}}}{\hbar}\left(1-\sqrt{\left|{\rm cos}\left(\pi\dfrac{M \Delta i_{\rm pc}}{\Phi_{0}}\right)\right|}\right)t_{\rm z}.
\end{equation}
\noindent
The evolution operator $\widetilde{U}_{\rm dz}$ turns to be
\begin{equation}
\begin{split}
&\widetilde{U}_{\rm dz} = \\
&
\left(
    \begin{matrix}
        1 & 0 \\
        0 & \exp\left({\mathrm{i}\dfrac{4\sqrt{E_{\rm C}E_{\rm J}}}{\hbar}\left(1-\sqrt{\left|{\rm cos}\left(\pi\dfrac{M \Delta i_{\rm pc}}{\Phi_{0}}\right)\right|}\right)t_{\rm z}}\right)
    \end{matrix}
\right).
\end{split}
\end{equation}

\par
By designing the qubit and the QET, the parameters $E_{\rm C}$, $E_{\rm J}$, $M$ and $\Delta i_{\rm pc}$ can be determined properly to make $t_{\rm z}$ in a range easy to realize. Then, for more precise control, the value of $t_{\rm z}$ should be optimized in practical experiments. Fine tuning can also be performed to compensate gate errors. In this simulation as a simple case, there are $E_{\rm J}/ \hbar=2\pi\cdot (11.147\ {\rm GHz})$, $E_{\rm C}/ \hbar=2\pi\cdot (148.628\ {\rm MHz})$, $M=0.02\ {\rm nH}$, $\Delta i_{\rm pc}=13.58935\ {\rm \mu A}$. And to realize a Z gate, $t_{\rm z}$ should be 2.261 ns by solving the equation $\varphi = \pi$. The initial state of Qubit 1 is set to be $|\psi\rangle_{\rm init}=1/\sqrt{2}|0\rangle+1/\sqrt{2}|1\rangle$. The data of $i_{\rm p}(t)$ is first extracted from its time-domain simulation in WRSpice similar with the former simulation of the single QET without the qubit. Then it is imported to the Z gate simulation program using QuTip \cite{johansson2012qutip, johansson2013qutip2} to calculate the time-domain data of the drive Hamiltonian for Z control. By calling the function solving master equation or Schrödinger equation of QuTip, like {\texttt {qutip.mesolve()}} or {\texttt {qutip.sesolve()}}, the time-evolution of the qubit state changed by a Z gate operated by the QET can be figured out with a total Hamiltonian $H$ consisting of drive Hamiltonian $H_{\rm dz}$ and idle qubit Hamiltonian $H_0$, which is expressed by Equation (\ref{HEqH0plusHdz}).
\par
The simulation results for a Z gate by the QET in the rotating frame is presented in FIG.~\ref{ZgateSimulationResult}. Besides the waveforms including (a) drive currents for DC/SFQ, (b) node voltages for QET input and (c) inductor loop current, (d) the frequencies of the qubit eigen energies and (e) the qubit frequency are also plotted in FIG.~\ref{ZgateSimulationResult}. The black trajectory of the point representing qubit state on the surface of Bloch sphere is drawn in FIG.~\ref{ZgateSimulationResultBlochSphere}. In this simulation, the gate operation time 2.261 ns is actually controlled by setting the time interval of rising edges of two square-wave pulses in FIG.~\ref{ZgateSimulationResult}(a). During the period of gate operation (in the lug boss of inductor loop current curve), the qubit frequency is kept at 4.779 GHz with $E_{\rm JS}/E_{\rm C}=137.5$, changed from 5.0 GHz with $E_{\rm JS}/E_{\rm C}=150$. And the end state of the qubit turns to be $|\psi\rangle_{\rm end}=0.70943|0\rangle+(-0.70476-0.0049432{\rm i})|1\rangle$, which is close to the ideal end state $|\psi\rangle_{\rm iend}=1/\sqrt{2}|0\rangle-1/\sqrt{2}|1\rangle$. The Z gate fidelity for only this time of operation is 99.99884\%.

\subsection*{C. $\rm{iSWAP}$ Gate by a QET}
\begin{figure*}[htb]
\begin{center}
\includegraphics[scale=0.7]{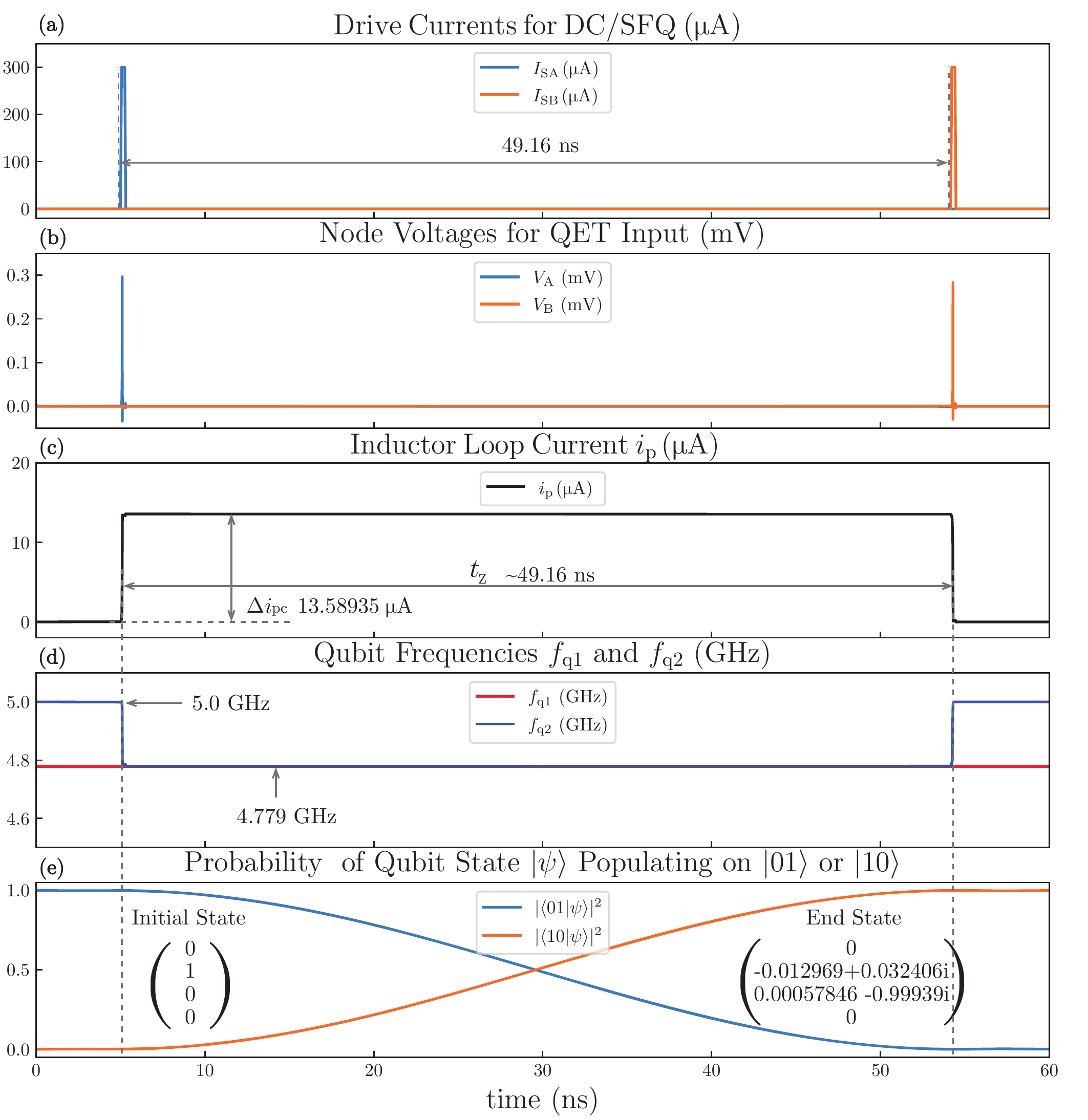}
\caption{\label{iswapSimulationResult} iSWAP Gate Simulation Result.}
\end{center}
\end{figure*}
\par
The simulation in this section shows how a QET can perform an iSWAP gate by making the frequency of Qubit 1 the same as that of Qubit 2. Compared with the former simulation for Z gate, the method of this simulation with WRSpice and QuTip remains unchanged, but the simulation circuit is enlarged, as shown in the green-dashed-line box in FIG.~\ref{qet_dcsfq_simcircuit}. The coupling strength between Qubit 1 and Qubit 2 is $g=2\pi\cdot (5\ {\rm MHz})$. The results is presented in FIG.~\ref{iswapSimulationResult}. The frequency of Qubit 1 (5.0 GHz) is tuned to the same level as the frequency of Qubit 2 (4.779 GHz) with $i_{\rm p}=\Delta i_{\rm pc}=13.58935\ {\rm \mu A}$ by inputting an SFQ pulse to port A of the QET. Then, the QET does nothing for $t_{\rm z}=49.16\ {\rm ns}$ to wait for state swaping between Qubit 1 and Qubit 2 with the initial state $|\psi\rangle_{\rm init}=|01\rangle$. When they finished swaping qubit state, the second SFQ pulse is inputted to port B of the QET, which makes Qubit 1 back to its idle frequency. Usually, for two qubits coupling with $g=2\pi\cdot (5\ {\rm MHz})$, the iSWAP gate needs 50 ns. However, here the gate operation time is optimized as 49.16 ns to eliminate the extra phase shift of Qubit 1 caused by changing its frequency and to ensure that the fidelity of the iSWAP gate is high enough at the same time. With the ideal end state $|\psi\rangle_{\rm iend}=|10\rangle$ and the actual end state $|\psi\rangle_{\rm end}=(-0.012969+0.032406{\rm i})|01\rangle+(0.00057846-0.99939{\rm i})|10\rangle$ in the simulation, the fidelity of the iSWAP gate for only this time of operation is 99.93906\%.

\section{Summary and Outlook}
\par
In conclusion, we have proposed a device called qubit energy tuner (QET) with description of its circuit structure and theory for its SFQ-based digital Z control to a flux-tunable transmon. A QET can convert SFQ pulses to external flux for qubits, so it is able to set the idle frequency of a flux-tunable transmon and at the same time perform gate operations involving Z control, like Z gates and iSWAP gates, thus paving an approach for digital Z control of an SFQ-based quantum-classical interface, which is highly desirable for the research and development of a large-scale superconducting quantum computer.
\par
For integrating with flux-tunable transmons and avoiding noise from SFQ circuits simultaneously, the parts of QETs consisting of flux bias units can be fabricated on another substrate which is electrical connected to the qubit chip through through silicon vias (TSVs) and indium bumps of a silicon interposer \cite{rosenberg20173d, rosenberg2020solid}. To realize mutual inductances between transmon SQUIDs and inductor loops of QETs, TSVs and indium bumps should be parts of inductor loops so that the piece of inductor line of an inductor loop for flux bias can be fabricated on the qubit chip or the surface of the silicon interposer faced to qubits. To eliminate the electrical loss of inductor loops, the material of TSVs should be superconductive, e.g. TiN. QETs may also be used in other application scenarios requiring flux tuning, such as CZ gates \cite{xu2020high}, flux-tunable couplers \cite{chen2014qubit, yan2018tunable} and qubit readout with a Josephson photomultiplier \cite{opremcak2018measurement}. As for further research, it is valuable to design and fabricate this device for experiments about SFQ-based digital control of qubits, especially flux-tunable transmons.

\section*{Acknowlegdments}
This work is partially supported by the key R$\&$D program of Guangdong province (Grant No.2019B010143002).

\section*{Appendix A. External Flux of SQUID from the Inductor Loop of QET}

\par
According to Kirchhoff's voltage law, the electric potentials of nodes A, B, C, D, E in FIG.~\ref{qet} are
\begin{equation}
V_{\rm A}(t^{\prime})=L_{1}\dfrac{{\rm d}i_{1}(t^{\prime})}{{\rm d}t^{\prime}}+M_{1}\dfrac{{\rm d}i_{\rm p}(t^{\prime})}{{\rm d}t^{\prime}}+M_{12}\dfrac{{\rm d}i_{2}(t^{\prime})}{{\rm d}t^{\prime}}, \label{voltA}
\end{equation}

\begin{equation}
V_{\rm B}(t^{\prime})=L_{2}\dfrac{{\rm d}i_{2}(t^{\prime})}{{\rm d}t^{\prime}}-M_{2}\dfrac{{\rm d}i_{\rm p}(t^{\prime})}{{\rm d}t^{\prime}}+M_{12}\dfrac{{\rm d}i_{1}(t^{\prime})}{{\rm d}t^{\prime}}, \label{voltB}
\end{equation}

\begin{equation}
V_{\rm C}(t^{\prime})=L_{3}\dfrac{{\rm d}i_{3}(t^{\prime})}{{\rm d}t^{\prime}}+M_{3}\dfrac{{\rm d}i_{\rm p}(t^{\prime})}{{\rm d}t^{\prime}}+M_{34}\dfrac{{\rm d}i_{4}(t^{\prime})}{{\rm d}t^{\prime}}, \label{voltC}
\end{equation}

\begin{equation}
V_{\rm D}(t^{\prime})=L_{4}\dfrac{{\rm d}i_{4}(t^{\prime})}{{\rm d}t^{\prime}}-M_{4}\dfrac{{\rm d}i_{\rm p}(t^{\prime})}{{\rm d}t^{\prime}}+M_{34}\dfrac{{\rm d}i_{3}(t^{\prime})}{{\rm d}t^{\prime}}, \label{voltD}
\end{equation}

\begin{equation}
\begin{split}
V_{\rm{E}}(t^{\prime})&=L_{\Sigma}\dfrac{{\rm d}i_{\rm p}(t^{\prime})}{{\rm d}t^{\prime}}+M_{1}\dfrac{{\rm d}i_{1}(t^{\prime})}{{\rm d}t^{\prime}}-M_{2}\dfrac{{\rm d}i_{2}(t^{\prime})}{{\rm d}t^{\prime}}\\
&+M_{3}\dfrac{{\rm d}i_{3}(t^{\prime})}{{\rm d}t^{\prime}}-M_{4}\dfrac{{\rm d}i_{4}(t^{\prime})}{{\rm d}t^{\prime}}+M\dfrac{{\rm d}i_{\rm q}(t^{\prime})}{{\rm d}t^{\prime}}, \label{voltE}
\end{split}
\end{equation}

\noindent
where

\begin{equation}
L_{\Sigma}=L_{\rm n0}+L_{\rm n1}+L_{\rm n2}+L_{\rm n3}+L_{\rm n4}+L_{\rm n5} \label{totalL}
\end{equation}

\noindent
is the total inductance of the inductor loop obtained by summing self inductances of all parts of the inductor loop. $L_{1}$, $L_{2}$, $L_{3}$ and $L_{4}$ are self inductances of inductors in flux bias units. $M_{1}$, $M_{2}$, $M_{3}$ and $M_{4}$ are mutual inductances of flux bias units and the inductor loop as shown in FIG.~\ref{qet}. $M$ is the mutual inductance between the inductor loop and the SQUID of flux-tunable transmon. $i_{1}(t^{\prime})$, $i_{2}(t^{\prime})$, $i_{3}(t^{\prime})$, $i_{4}(t^{\prime})$ are currents of inductors in flux bias units at the moment $t^{\prime}$. $i_{\rm p}(t^{\prime})$ and $i_{\rm q}(t^{\prime})$  are the current of inductor loop and the SQUID at the moment $t^{\prime}$. 
\par
Under the zero initial condition, integrating both sides of Equation (\ref{voltA})-(\ref{voltE}) with 0 as lower bound and time $t$ as upper bound yields

\begin{equation}
\int^{t}_{0}V_{\rm A}(t^{\prime}){\rm d}t^{\prime}=L_{1}i_{1}(t)+M_{1}i_{\rm p}(t)+M_{12}i_{2}(t), \label{PhiA}
\end{equation}

\begin{equation}
\int^{t}_{0}V_{\rm B}(t^{\prime}){\rm d}t^{\prime}=L_{2}i_{2}(t)-M_{2}i_{\rm p}(t)+M_{12}i_{1}(t), \label{PhiB}
\end{equation}

\begin{equation}
\int^{t}_{0}V_{\rm C}(t^{\prime}){\rm d}t^{\prime}=L_{3}i_{3}(t)+M_{3}i_{\rm p}(t)+M_{34}i_{4}(t), \label{PhiC}
\end{equation}

\begin{equation}
\int^{t}_{0}V_{\rm D}(t^{\prime}){\rm d}t^{\prime}=L_{4}i_{4}(t)-M_{4}i_{\rm p}(t)+M_{34}i_{3}(t), \label{PhiD}
\end{equation}

\begin{equation}
\begin{split}
\int^{t}_{0}V_{\rm{E}}(t^{\prime}){\rm d}t^{\prime}&=L_{\Sigma}i_{\rm p}(t)+M_{1}i_{1}(t)-M_{2}i_{2}(t)\\
&+M_{3}i_{3}(t)-M_{4}i_{4}(t)+Mi_{\rm q}(t). \label{PhiE}
\end{split}
\end{equation}
The mutual inductance $M$ is designed to be much smaller than total inductance of the inductor loop $L_{\Sigma}$ and other mutual inductances like $M_{1}$ for weak coupling to the SQUID of the qubit. And the ring current of the SQUID $i_{\rm q}$(t) should be less than the critical current of its Josephson junctions, which is about tens of nA for Al/AlOx/Al junctions and smaller than the current in the inductance loop $i_{\rm p}$(t) (about several or tens of mA) by two or more orders of magnitude. Therefore, the influence of the SQUID on the inductance loop, $Mi_{\rm q}$, can be ignored in Equation (\ref{PhiE}), and the electric potential of node E is rewritten as
\begin{equation}
\begin{split}
\int^{t}_{0}V_{\rm{E}}(t^{\prime}){\rm d}t^{\prime}&=L_{\Sigma}i_{\rm p}(t)+M_{1}i_{1}(t)-M_{2}i_{2}(t)\\
&+M_{3}i_{3}(t)-M_{4}i_{4}(t). \label{PhiErewritten}
\end{split}
\end{equation}

\par
And then, let
\begin{equation}
\varPhi_{\rm A}(t)=\int^{t}_{0}V_{\rm A}(t^{\prime}){\rm d}t^{\prime}, \label{varPhiA}
\end{equation}

\begin{equation}
\varPhi_{\rm B}(t)=\int^{t}_{0}V_{\rm B}(t^{\prime}){\rm d}t^{\prime}, \label{varPhiB}
\end{equation}

\begin{equation}
\varPhi_{\rm C}(t)=\int^{t}_{0}V_{\rm C}(t^{\prime}){\rm d}t^{\prime}, \label{varPhiC}
\end{equation}

\begin{equation}
\varPhi_{\rm D}(t)=\int^{t}_{0}V_{\rm D}(t^{\prime}){\rm d}t^{\prime}, \label{varPhiD}
\end{equation}

\begin{equation}
\varPhi_{\rm E}(t)=\int^{t}_{0}V_{\rm E}(t^{\prime}){\rm d}t^{\prime}, \label{varPhiE}
\end{equation}

\noindent
we get
\begin{equation}
{\bm \varPhi}(t)={\bm L}{\bm i}(t), 
\end{equation}
\noindent
where
\begin{equation}
{\bm \varPhi}(t)=
\begin{bmatrix}
\varPhi_{\rm A}(t) \\
\varPhi_{\rm B}(t)\\
\varPhi_{\rm C}(t) \\
\varPhi_{\rm D}(t)\\
\varPhi_{\rm E}(t)
\end{bmatrix}
,
\end{equation}

\begin{equation}
{\bm i}(t)=
\begin{bmatrix}
i_{1}(t) \\
i_{2}(t) \\
i_{3}(t) \\
i_{4}(t) \\
i_{\rm p}(t)
\end{bmatrix}
,
\end{equation}
and 
\begin{equation}
{\bm L}=
\begin{bmatrix}
L_{1} &M_{12} &0 &0 &M_{1} \\
M_{12}  &L_{2} &0 &0 &-M_{2} \\
0 &0 &L_{3} &M_{34} &M_{3} \\
0 &0 &M_{34} &L_{4} &-M_{4} \\
M_{1} &-M_{2} &M_{3} &-M_{4} &L_{\Sigma}
\end{bmatrix}
.
\end{equation}

\par
Then, to get ${\bm i}(t)$, we have

\begin{equation}
{\bm i}(t)={\bm L^{-1}}{\bm \varPhi}(t),
\end{equation}
\noindent
where ${\bm L^{-1}}=\dfrac{1}{F}{\bm A}$. $\dfrac{1}{F}$ is the common factor of the elements in the inverse matrix of ${\bm L}$. $F$ is
\begin{equation}
\begin{split}
F&=L_{2}(M_{34}^{2}(L_{1}L_{\Sigma}-M_{1}^{2})+2L_{1}M_{3}M_{34}M_{4}\\
&+L_{3}(-L_{4}(L_{1}L_{\Sigma}-M_{1}^{2})+L_{1}M_{4}^{2})+L_{1}L_{4}M_{3}^{2})\\
&+(-L_{1}M_{2}^{2}-L_{\Sigma}M_{12}^{2}-2M_{1}M_{12}M_{2})M_{34}^{2}\\
&-2M_{12}^{2}M_{3}M_{34}M_{4}\\
&+L_{3}((L_{1}M_{2}^{2}+L_{\Sigma}M_{12}^{2}+2M_{1}M_{12}M_{2})L_{4}-M_{12}^{2}M_{4}^{2})\\
&-L_{4}M_{12}^{2}M_{3}^{2}.
\end{split}
\end{equation}

The elements $a_{ij}$ ($i,j=1,2,3,4,5$) of ${\bm A}$, are
\begin{subequations}
\begin{eqnarray*}
a_{11}&=&L_{2}(M_{34}^{2}L_{\Sigma}+2M_{3}M_{4}M_{34}\\
&&+(-L_{4}L_{\Sigma}+M_{4}^{2})L_{3}+L_{4}M_{3}^{2})\\
&&+M_{2}^{2}(L_{3}L_{4}-M_{34}^{2}),\\
a_{12}&=&M_{12}(-M_{34}^{2}L_{\Sigma}-2M_{3}M_{4}M_{34}\\
&&+(L_{4}L_{\Sigma}-M_{4}^{2})L_{3}-L_{4}M_{3}^{2})\\
&&+M_{1}M_{2}(L_{3}L_{4}-M_{34}^{2}),\\
a_{13}&=& -(L_{4}M_{3}+M_{34}M_{4})(L_{2}M_{1}+M_{12}M_{2}),\\
a_{14}&=&(L_{2}M_{1}+M_{12}M_{2})(L_{3}M_{4}+M_{3}M_{34}),\\
a_{15}&=&(L_{3}L_{4}-M_{34}^{2})(L_{2}M_{1}+M_{12}M_{2}),\\
&&\\
\end{eqnarray*}
\end{subequations}
\begin{subequations}
\begin{eqnarray*}
a_{21}&=&M_{12}(-M_{34}^{2}L_{\Sigma}-2M_{3}M_{4}M_{34}\\
&&+(L_{4}L_{\Sigma}-M_{4}^{2})L_{3}-L_{4}M_{3}^{2})\\
&&+M_{1}M_{2}(L_{3}L_{4}-M_{34}^{2}),\\
a_{22}&=&L_{1}(M_{34}^{2}L_{\Sigma}+2M_{3}M_{4}M_{34}\\
&&+(-L_{4}L_{\Sigma}+M_{4}^{2})L_{3}+L_{4}M_{3}^{2})\\
&&+M_{1}^{2}(L_{3}L_{4}-M_{34}^{2}),\\
a_{23}&=&(L_{4}M_{3}+M_{34}M_{4})(L_{1}M_{2}+M_{1}M_{12}),\\
a_{24}&=&-(L_{1}M_{2}+M_{1}M_{12})(L_{3}M_{4}+M_{3}M_{34}),\\
a_{25}&=&-(L_{3}L_{4}-M_{34}^{2})(L_{1}M_{2}+M_{1}M_{12}),\\
&&\\
a_{31}&=&-(L_{4}M_{3}+M_{34}M_{4})(L_{2}M_{1}+M_{12}M_{2}),\\
a_{32}&=&(L_{4}M_{3}+M_{34}M_{4})(L_{1}M_{2}+M_{1}M_{12}),\\
a_{33}&=&L_{4}(L_{\Sigma}M_{12}^{2}+2M_{1}M_{2}M_{12}\\
&&+(-L_{2}L_{\Sigma}+M_{2}^{2})L_{1}+L_{2}M_{1}^{2})\\
&&+M_{4}^{2}(L_{1}L_{2}-M_{12}^{2}),\\
a_{34}&=&M_{34}(-L_{\Sigma}M_{12}^{2}-2M_{1}M_{2}M_{12}\\
&&+(L_{2}L_{\Sigma}-M_{2}^{2})L_{1}-L_{2}M_{1}^{2})\\
&&+M_{3}M_{4}(L_{1}L_{2}-M_{12}^{2}),\\
a_{35}&=&(L_{1}L_{2}-M_{12}^{2})(L_{4}M_{3}+M_{34}M_{4}),\\
&&\\
a_{41}&=&(L_{2}M_{1}+M_{12}M_{2})(L_{3}M_{4}+M_{3}M_{34}),\\
a_{42}&=&-(L_{1}M_{2}+M_{1}M_{12})(L_{3}M_{4}+M_{3}M_{34}),\\
a_{43}&=&M_{34}(-L_{\Sigma}M_{12}^{2}-2M_{1}M_{2}M_{12}\\
&&+(L_{2}L_{\Sigma}-M_{2}^{2})L_{1}-L_{2}M_{1}^{2})\\
&&+M_{3}M_{4}(L_{1}L_{2}-M_{12}^{2}),\\
a_{44}&=&L_{3}(L_{\Sigma}M_{12}^{2}+2M_{1}M_{2}M_{12}\\
&&+(-L_{2}L_{\Sigma}+M_{2}^{2})L_{1}+L_{2}M_{1}^{2})\\
&&+M_{3}^{2}(L_{1}L_{2}-M_{12}^{2}),\\
a_{45}&=&-(L_{1}L_{2}-M_{12}^{2})(L_{3}M_{4}+M_{3}M_{34}),\\
&&\\
a_{51}&=&(L_{3}L_{4}-M_{34}^{2})(L_{2}M_{1}+M_{12}M_{2}),\\
a_{52}&=&-(L_{3}L_{4}-M_{34}^{2})(L_{1}M_{2}+M_{1}M_{12}),\\
a_{53}&=&(L_{1}L_{2}-M_{12}^{2})(L_{4}M_{3}+M_{34}M_{4}),\\
a_{54}&=&-(L_{1}L_{2}-M_{12}^{2})(L_{3}M_{4}+M_{3}M_{34}),\\
a_{55}&=&-(L_{3}L_{4}-M_{34}^{2})(L_{1}L_{2}-M_{12}^{2}).
\end{eqnarray*}
\end{subequations}
\noindent
Therefore, we have

\begin{equation}
\begin{split}
i_{\rm p}(t)=&\dfrac{1}{F}(a_{51}\varPhi_{\rm A}(t)+a_{52}\varPhi_{\rm B}(t)\\
&+a_{53}\varPhi_{\rm C}(t)+a_{54}\varPhi_{\rm D}(t)+a_{55}\varPhi_{\rm E}(t)),
\end{split}
\end{equation}
\noindent
that is
\begin{equation}
\begin{split}
i_{\rm p}(t)=&\dfrac{1}{F}(\varPhi_{\rm A}(t)(L_{3}L_{4}-M_{34}^{2})(L_{2}M_{1}+M_{12}M_{2})\\
&-\varPhi_{\rm B}(t)(L_{3}L_{4}-M_{34}^{2})(L_{1}M_{2}+M_{1}M_{12})\\
&+\varPhi_{\rm C}(t)(L_{1}L_{2}-M_{12}^{2})(L_{4}M_{3}+M_{34}M_{4})\\
&-\varPhi_{\rm D}(t)(L_{1}L_{2}-M_{12}^{2})(L_{3}M_{4}+M_{3}M_{34})\\
&-\varPhi_{\rm E}(t)(L_{3}L_{4}-M_{34}^{2})(L_{1}L_{2}-M_{12}^{2})).
\end{split}
\end{equation}
\noindent
Because node E is connected to the ground, $\varPhi_{\rm E}$ should always be zero. To make the flux bias units of coarse tuning be able to increase or decrease the external flux through the SQUID by the same amount, the following requirements should be met: 
\begin{subequations}
\begin{eqnarray}
L_{1}&=&L_{2}=L_{\rm c},\\
M_{1}&=&M_{2}=M_{\rm c}.
\end{eqnarray}
\end{subequations}
Similarly, for the flux bias units of fine tuning, we have
\begin{subequations}
\begin{eqnarray}
L_{3}&=&L_{4}=L_{\rm f},\\
M_{3}&=&M_{4}=M_{\rm f}.
\end{eqnarray}
\end{subequations}

\noindent
Therefore, $i_{\rm p}(t)$ turns to be 

\begin{equation}
\begin{split}
i_{\rm p}(t)=&\dfrac{1}{F}((\varPhi_{\rm A}-\varPhi_{\rm B})(L_{\rm f}^{2}-M_{34}^{2})(L_{\rm c}M_{\rm c}+M_{12}M_{\rm c})\\
&+(\varPhi_{\rm C}-\varPhi_{\rm D})(L_{\rm c}^{2}-M_{12}^{2})(L_{\rm f}M_{\rm f}+M_{34}M_{\rm f})), \label{ipt2}
\end{split}
\end{equation}

\noindent
and $F$ turns to be

\begin{equation}
\begin{split}
F&=L_{\rm c}L_{\rm f}((-L_{\Sigma}L_{\rm f}+M_{\rm f}^2)L_{\rm c}+L_{\rm f}M_{\rm c}^2)\\
&+(L_{\rm f}+M_{34})(L_{\Sigma}L_{\rm f}-L_{\Sigma}M_{34}-2M_{\rm f}^2)M_{12}^2\\
&+(2L_{\rm f}^2M_{\rm c}^2-2M_{34}^2M_{\rm c}^2)M_{12}\\
&+L_{\rm c}((L_{\Sigma}L_{\rm c}-2M_{\rm c}^2)M_{34}^2\\
&+2L_{\rm c}M_{\rm f}M_{\rm f}M_{34}+L_{\rm c}L_{\rm f}M_{\rm f}^2+L_{\rm f}^2M_{\rm c}^2).
\end{split}
\end{equation}
\noindent
Hence, the relationship between the current of the inductor loop $i_{\rm p}(t)$ and the external flux through the SQUID $\varPhi_{\rm e}$ is 
\begin{equation}
\varPhi_{\rm e} = Mi_{\rm p}(t).
\end{equation}

\bibliography{reference}

\end{document}